# Some new considerations about the \*nu*-function

**Dušan POPOV**

**University Politehnica Timisoara, Department of Physical Foundations of Engineering,
Bd. V. Parvan No.2, 300223 Timisoara, Romania**
dusan_popov@yahoo.co.uk
**ORCID:** https://orcid.org/0000-0003-3631-3247

**Abstract**

The present paper starts from a previously deduced result, in which the \*nu*-function plays the role of the normalization function of generalized hypergeometric coherent states for quantum systems with a continuous spectrum. We have generalized this idea, obtaining a new function – the generalized \*nu*-function. By defining a discrete-continuous limit, we revealed a series of interesting properties that, in the last instance, allow the formulation and solution of new integrals involving the generalized \*nu*-functions which depend on both scalar arguments as well as those containing creation and annihilation operators, which generate the generalized hypergeometric coherent states. To our knowledge, the results obtained by us do not appear in the literature.

## 1. Introduction

As is well known, for some function $f(E)$, the Laplace transform is defined by the integral

$$\mathcal{L}\{f\}(s) = \int_0^\infty dE\, e^{-sE} f(E) \tag{1.1}$$

where $s$ is a complex number.

On the other hand, the reciprocal gamma function is

$$f(E+1) = \frac{1}{\Gamma(E+1)} \tag{1.2}$$

By combining these two relations, we obtain

$$\mathcal{L}\{f\}(e^{-s}) = \int_0^\infty dE\, \frac{(e^{-s})^E}{\Gamma(E+1)} \tag{1.3}$$

Considering $e^{-s} = z$, where $z$ is complex number $z = |z|\exp(i\varphi)$, $0 \le |z| \le R_c \le \infty$, and $R_c$ is the radius of convergence of the power series in the variable $|z|$, we will get

$$\mathcal{L}\{f\}(z) = \int_0^\infty dE\, \frac{z^E}{\Gamma(E+1)} \equiv \nu(z) \tag{1.4}$$

A new function $\nu(z)$ has appeared on the right side, called *nu-function*, which is therefore a generalization of the Laplace transform of the reciprocal gamma function. It was introduced by Volterra in 1916 [1]. In the literature, apart from two important books [1] and [2], there are not many references to the *nu*-function $\nu(z)$.

Also, another function $\nu(z, \alpha)$ was defined as an extension or generalization of $\nu(z)$, through the relation [1]:

$$\nu(z, \alpha) = \int_0^\infty dE\, \frac{z^{\alpha+E}}{\Gamma(\alpha+E+1)} \tag{1.5}$$

The relationship between these two functions is

$$\frac{d^n}{dz^n}\nu(z) = \nu(z, -n) \tag{1.6}$$

On the other hand, the most general form of coherent states (CSs) $|z>$ from quantum mechanics, expanded in the Fock-vector's basis $\{|n>, n = 0,1,2,...,n_{\max} \le \infty\}$, which are called the generalized hypergeometric coherent states (GH-CSs) have the expression [3]:

$$|z> = \frac{1}{\sqrt{{}_pF_q\left(\{a_i\}_1^p;\{b_j\}_1^q;|z|^2\right)}} \sum_{n=0}^{n_{\max} \le \infty} \frac{z^n}{\sqrt{\rho_{p,q}(n)}} |n> \tag{1.7}$$

Here $p$ and $q$ are real integers, $(a_i)_n$ are the Pochhammer's symbols, and we will use the abbreviated notations $x_1, x_2, ..., x_m \equiv \{x_i\}_1^m$.

The name "generalized hypergeometric coherent states" comes from the fact that their normalizing function ${}_pF_q\left(\{a_i\}_1^p;\{b_j\}_1^q;|z|^2\right)$ is a generalized hypergeometric function. At the same time, the positive constants $\rho_{p,q}(n)$ are assumed to arise as the moments of a probability distribution [4] and for GH-CSs they are defined as follows [5]:

$$\rho_{p,q}(n) \equiv n!\frac{\prod_{j=1}^{q}(b_j)_n}{\prod_{i=1}^{p}(a_i)_n} \tag{1.8}$$

Consequently, the normalization functions, i.e. the generalized hypergeometric functions are connected with the structure constants as [6]

$$_pF_q\left(\{a\}_1^p;\{b\}_1^q;|z|^2\right)=\sum_{n=0}^{\infty}\frac{\prod_{i=1}^{p}(a_i)_n}{\prod_{j=1}^{q}(b_j)_n}\frac{\left(|z|^2\right)^n}{n!}=\sum_{n=0}^{\infty}\frac{1}{\rho_{p,q}(n)}\left(|z|^2\right)^n \tag{1.9}$$

It is known that any coherent state must satisfy some minimal conditions, generically called "Klauder's prescriptions" [7]:

(I). They must be normalized but non-orthogonal, and to form an overcomplete set.

$$<z|z'> = \begin{cases} 1, & z=z' \\ \neq 0, & z \neq z' \end{cases} \tag{1.10}$$

(II). They must be continuous in the label variable $z$:

$$\lim_{z \to z'}\|z-z'\| = \lim_{\substack{|z|\to|z'| \\ \varphi \to \varphi'}}\sqrt{|z|^2+|z'|^2-2|z||z'|\cos(\varphi-\varphi')} = 0 \tag{1.11}$$

(III). It must necessarily satisfy the resolution of the identity operator, i.e. to close a resolution of the identity (this relation is called the completeness relation):

$$\int d\mu(z)|z><z| = \hat{I} = \sum_{n=0}^{\infty}|n><n| \tag{1.12}$$

with the integration measure $d\mu(z) = \frac{d^2z}{\pi} = \frac{d\varphi}{2\pi}d\left(|z|^2\right)h\left(|z|^2\right)$ and positive weight function $h\left(|z|^2\right)$ which must be found for each individual kind of CSs.

Although at first glance, there seems to be any connection between the two previously defined concepts / entities, i.e. *nu*-function $v(z)$ and generalized coherent states (GH-CSs) $|z>$, in the paper [8] we still found such an interesting connection which can be considered a first application of *nu*-function. In the present paper, we propose to deepen this connection, which will result in a series of new properties of the *nu*-function $v(z)$, which do not appear in the specialized literature.

In paper [8] we studied the transition from the *discontinuous spectrum* (*d*) to the *continuous spectrum* (*c*) of an certain quantum system. We found that if a certain limit, called *the discrete – continuous limit* $d \to c$, is applied, a quantity that characterizes a system with a discontinuous spectrum will pass, at this limit, into the corresponding quantity connected with the continuous spectrum. In quantum mechanics, there are systems that have both a discontinuous and a continuous spectrum. Perhaps the closest example is the diatomic molecule, whose inter nuclear potential is a Morse-type potential, characterized by discontinuous energy levels (corresponding to the so-called bound states). For higher values of the main quantum number $n$, i.e. for energies that exceed the dissociation energy $D_e$, due to external causes (e.g. temperature rise), the bound system of the two nuclei dissociates. The two nuclei (with the corresponding electrons) move like free particles, having a continuous energy spectrum.

Let us highlight here that, for a certain quantum system, two types of CSs can be defined: a) *Barut-Girardello type*, defined as eigenvectors of the annihilation operator [9], and b) *Klauder-Perelomov type*, defined as the result of the action of the displacement operator on the vacuum state [10]. Their expansions in the set of Fock vectors are different, but they are dual. The duality is manifested by the fact that the indices p and q, as well as the sets of numbers a and b are interchangeable [11]. However, if the discrete - continuous d-c limit is applied to them, then their expressions tend to the same mathematical form, similar to the coherent states of the one-dimensional harmonic oscillator (HO-1D). That's why, considering the purpose of the present paper, we will only focus, as a starting point, on the expressions of Barut-Girardello type coherent states.

## 2. The main results from [8] - in short

To begin with, let us we consider a dimensionless Hamiltonian $\mathcal{H}$ with a non-degenerate continuous spectrum, and dimensionless eigenstates $|E>$, $\hbar\omega=1$, (with $0 \le E \le \infty$) which are formal delta-function normalized, i.e., with $<E|E'>=\delta(E-E')$.

$$\mathcal{H}|E>=E|E> \tag{2.1}$$

The closure or completeness relation for continuous spectrum is

$$\int_0^\infty dE |E><E| = I , \quad \text{i.e} \quad \int_0^\infty dE <E'|E><E|E''> = \delta(E'-E'') \tag{2.2}$$

The coherent states of the *continuous spectrum* can be obtained if we use the following limit (for brevity we will call it *the discrete – continuous limit* $d \to c$), defined as (see, also, [12]):

$$X_c(E) = \lim_{\substack{n \to E \\ n_{max} \to \infty}} X_d(n, n_{max}) \equiv \lim_{d \to c} X_d(n, n_{max}), \quad \sum_{n=0}^{n_{max}} X_d(n, n_{max}) \to \int_0^\infty dE\, X_c(E)$$

$$\sum_{n=0}^\infty \to \int_0^\infty dE$$
$$p = q$$
$$\{a_i\} = \{b_j\} \tag{2.3}$$

So, the connection between the observables or quantities of the discontinuous spectrum $X_d$ and those corresponding to the continuous spectrum $X_c$ requires the following operations:

*a)* the energy quantum number $n$ must be replaced by the dimensionless energy $E$; *b)* the maximal number of bound states must tend to infinity $n_{max} \to \infty$; *c)* simultaneously, the sum with respect to $n$ must be replaced by the integral with respect to $E$; *d)* the indices $p$ and $q$ of the generalized hypergeometric functions, as well as the sets of parameters $\{a_i\}_1^p$ and $\{b_j\}_1^q$ must be equal.

Consequently, we obtain the following limits:

$$\lim_{d \to c} \rho_{p,q}(n) = \lim_{d \to c} n! \frac{\prod_{j=1}^{q}(b_j)_n}{\prod_{i=1}^{p}(a_i)_n} = \Gamma(E+1) \qquad (2.4)$$

$$\lim_{d \to c} {}_pF_q(\{a_i\}_1^p; \{b_j\}_1^q; |z|^2) = \lim_{d \to c} \sum_{n=0}^{\infty} \frac{1}{\rho_{p,q}(n)} (|z|^2)^n = \int_0^{\infty} dE \frac{(|z|^2)^E}{\Gamma(E+1)} = \nu(|z|^2) \qquad (2.5)$$

This leads to the following limit for GH-CSs:

$$|z> = \lim_{d \to c} |z> = \frac{1}{\sqrt{\nu(|z|^2)}} \int_0^{\infty} dE \frac{z^E}{\sqrt{\Gamma(E+1)}} |E> \qquad (2.6)$$

The expression of CSs for continuous spectrum was firstly obtained, *by another method and considerations*, for the Gazeau-Klauder CSs, in [4] and later in [13] and [14].

The overlap of two CSs follows immediately

$$<z|z'> = \lim_{d \to c} <z|z'> = \frac{\nu(z^* z')}{\sqrt{\nu(|z|^2)}\sqrt{\nu(|z'|^2)}} \qquad (2.7)$$

*Observation:* In order not to overload the writing of formulas, we kept the same variable $z$ for both the discontinuous and the continuous spectrum, in other words $\lim_{d \to c} z = z$. Only where necessary, in order not to create confusion, we will explain the indices $d$ or $c$.

Using the Mellin transform of the $G$-function [6]

$$\int_0^{\infty} dx\, x^{s-1} G_{p,q}^{m,n}\left(\omega x \left| \begin{array}{c} \{a_i\}_1^n; \{a_i\}_{n+1}^p \\ \{b_j\}_1^m; \{b_j\}_{m+1}^q \end{array} \right.\right) = \frac{1}{\omega^s} \frac{\sum_{j=1}^{m}\Gamma(b_j+s)\sum_{i=1}^{n}\Gamma(1-a_i-s)}{\sum_{j=m+1}^{q}\Gamma(1-b_j-s)\sum_{i=n+1}^{p}\Gamma(a_i+s)}, \quad \begin{array}{c} p<q \\ \omega \neq 0 \end{array} \qquad (2.8)$$

the integration measure of a GH-CSs (1.5) for discontinuous spectra was obtained in [15] and this is

$$d\mu_{p,q}^d(z) = \frac{d\varphi}{2\pi} d(|z|^2) \frac{\prod_{i=1}^{p}\Gamma(a_i)}{\prod_{j=1}^{q}\Gamma(b_j)} {}_pF_q(\{a_i\}_1^p; \{b_j\}_1^q; |z|^2) G_{p,q+1}^{q+1,0}\left(|z|^2 \left| \begin{array}{c} /; \{a_i-1\}_1^p \\ 0, \{b_j-1\}_1^q; / \end{array}\right.\right) \qquad (2.9)$$

so that its limit is

$$\lim_{d \to c} d\mu_{p,q}^d(z) = \frac{d\varphi}{2\pi} d(|z|^2) e^{-|z|^2} \nu(|z|^2) \equiv d\mu^c(z) \qquad (2.10)$$

where we took into account the specialized values of the Meijer's G functions $G_{0,1}^{1,0}(|z|^2|0) = \exp(-|z|^2)$ [6].

In this context, the following relationship is also valid

$$\lim_{d \to c} \int d\mu_{p,q}^d(z) |z><z| = \int d\mu^c(z) |z><z| = \int_0^\infty dE |E><E| = \hat{I} \qquad (2.11)$$

In performing the demonstration through the corresponding substitutions, after the angular integration, we used a *fundamental integral*:

$$\int_0^\infty d(|z|^2)(|z|^2)^E G_{p,q+1}^{q+1,0}\left(|z|^2 \Big|_{0, \{b_j - 1\}_1^q;}^{/; \quad \{a_i - 1\}_1^p}\right) = \Gamma(E+1) \frac{\prod_{j=1}^q \Gamma(b_j + E)}{\prod_{i=1}^p \Gamma(a_i + E)} =$$

$$= \frac{\prod_{j=1}^q \Gamma(b_j)}{\prod_{i=1}^p \Gamma(a_i)} \rho_{p,q}(E) \qquad (2.12)$$

which will be useful for the following.

As we showed previously [5], the GH-CSs are generated by a pair of Hermitian creation / annihilation operators $\mathcal{A}_+$ and $\mathcal{A}_-$. A new operational ordering approach, called DOOT (the *diagonal ordering operation technique*), can be applied to them, which leads to a series of new results both for the discontinuous spectrum [15] and also for the continuous one [8].

The pair operators $\mathcal{A}_-$ and $\mathcal{A}_+$ are Hermitian $(\mathcal{A}_-)^+ = \mathcal{A}_+$ and satisfy the following equations:

$$\mathcal{A}_- |n> = \sqrt{n} |n-1> , \quad \mathcal{A}_+ |n> = \sqrt{(n+1)} |n+1> , \quad \mathcal{A}_+ \mathcal{A}_- |n> = n|n> \qquad (2.13)$$

Their action on the vectors $|E>$ results from the application of the *discrete – continuous limit* $d \to c$ on their discrete counterparts:

$$\mathcal{A}_- |E> = \sqrt{E} |E-1> , \quad \mathcal{A}_+ |E> = \sqrt{(E+1)} |E+1> , \quad \mathcal{A}_+ \mathcal{A}_- |E> = E|E> \qquad (2.14)$$

In [8] for the continuous spectrum we introduced a real dimensionless energy parameter $\varepsilon > 0$, which is *not a quanta*, and can be interpreted as a suitable "jump unity" in the energy scale of continuous spectra. By equating to unity $\varepsilon = 1$, the system's energy may be written simply as $E = m$. If we successively apply $m$-fold the raising operator $\mathcal{A}_+$ to the ground or vacuum state $|0>$, for the continuous spectrum we obtain

$$|E> = \frac{1}{\sqrt{\Gamma(E+1)}} (\mathcal{A}_+)^E |0> , \quad <E| = \frac{1}{\sqrt{\Gamma(E+1)}} <0|(\mathcal{A}_-)^E \qquad (2.15)$$

Starting from Eq. (2.2) and using the DOOT rules (for details, see [12], [15]), we will obtain the expression of the projector of the vacuum state from the continuous spectrum $|0><0|$:

$$\int_0^\infty dE |E><E| = |0><0| \int_0^\infty dE \frac{\#(\mathcal{A}_+ \mathcal{A}_-)^E \#}{\Gamma(E+1)} = |0><0| \#\nu(\mathcal{A}_+ \mathcal{A}_-)\# = 1 \qquad (2.16)$$

$$|0><0| = \frac{1}{\#\nu(\mathcal{A}_+ \mathcal{A}_-)\#} \qquad (2.17)$$

The sign #...# indicates a normal ordering of the operators, in the sense of the DOOT approach.

Consequently, the projector in the energy eigenvectors space $|E><E|$ is

$$|E><E| = \# \frac{1}{\nu(\mathcal{A}_+\mathcal{A}_-)} \frac{(\mathcal{A}_+\mathcal{A}_-)^E}{\Gamma(E+1)} \# \qquad (2.18)$$

With the above relations and the DOOT rules, the CSs for the continuous spectrum becomes:

$$|z> = \frac{1}{\sqrt{\nu(|z|^2)}} \int_0^\infty dE \frac{(z\mathcal{A}_+)^E}{\Gamma(E+1)}|0> = \frac{1}{\sqrt{\nu(|z|^2)}} \nu(z\mathcal{A}_+)|0> \qquad (2.19)$$

and similarly their counterpart, so that the projector on the CSs state $|z>$ is

$$|z><z| = \frac{1}{\nu(|z|^2)} \# \frac{\nu(z\mathcal{A}_+)\nu(z^*\mathcal{A}_-)}{\nu(\mathcal{A}_+\mathcal{A}_-)} \# \qquad (2.20)$$

The verification of this expression can be done starting from the completeness relation of CSs, Eq. (2.10).

$$\int d\mu^c(z) |z><z| = \# \frac{1}{\nu(\mathcal{A}_+\mathcal{A}_-)} \# \int_0^\infty d(|z|^2) e^{-|z|^2} \int_0^{2\pi} \frac{d\varphi}{2\pi} \# \nu(z\mathcal{A}_+)\nu(z^*\mathcal{A}_-)\# = 1 \qquad (2.21)$$

because the angular integral is

$$\int_0^{2\pi} \frac{d\varphi}{2\pi} \# \nu(z\mathcal{A}_+)\nu(z^*\mathcal{A}_-)\# = \int_0^\infty dE \, \# \frac{(\mathcal{A}_+\mathcal{A}_-)^E \#}{[\Gamma(E+1)]^2} (|z|^2)^E \qquad (2.22)$$

On the other hand, the probability density of the transition from state $|E>$ to state $|z>$ has the expression

$$P_E(|z|^2) \equiv |<z|E>|^2 = \frac{1}{\nu(|z|^2)} \frac{(|z|^2)^E}{\Gamma(E+1)} \qquad (2.23)$$

If we apply the *inverse* limit, i.e. the *continuous – discrete limit* $c \to d$, we obtain just the Poisson probability density function for the discrete case

$$\lim_{c \to d} P_E(|z|^2) \equiv \lim_{\substack{E \to n \\ \int \to \Sigma}} P_E(|z|^2) = e^{-|z|^2} \frac{(|z|^2)^n}{n!} \equiv P_n(|z|^2) \qquad (2.24)$$

From the relations above it can be seen that the *nu*-function $\nu(|z|^2)$ does find its place in the description of the continuous spectrum of a quantum system. To our knowledge, this is *the first application* of the *nu*-function $\nu(|z|^2)$ in a non-mathematical scientific field.

### 3. The generalized discrete – continuous limit $d \to c$

Let us now generalize the main results obtained in [8] and to examine what practical consequences this generalization has. Compared to the limit used in this paper, Eq. (2.3), in the next we will adopt a less restrictive limit:

$$X_c(E) = \lim_{\substack{n \to E \\ n_{max} \to \infty \\ \sum_{n=0}^{\infty} \to \int_0^{\infty} dE}} X_d(n, n_{max}) \equiv \lim_{d \to c} X_d(n, n_{max}) \ , \quad \sum_{n=0}^{n_{max}} X_d(n, n_{max}) \to \int_0^{\infty} dE\, X_c(E) \quad (3.1)$$

This means that all observables $X_c$ what characterizes the system with continuous spectrum will be obtained as a limiting case of the corresponding observables $X_d$ of the discrete spectrum, through three operations: *a*) replacing $n \to E$, by the dimensionless energy $E$; *b*) the extension $n_{max} \to \infty$; *c*) simultaneously, the sum with respect to $n$ must be replaced by the integral with respect to $E$. For this reason we will call this the *generalized discrete – continuous limit $d \to c$ limit* (Gd-cL). We will also introduce the notations for generalized hypergeometric coherent states: d-GH-CSs for discontinuous spectrum and c-GH-CSs for continuous.

Let's apply the Gd-cL on the elements characteristic for the discontinuous spectrum.

$$\lim_{d \to c} \rho_{p,q}(n) = \lim_{d \to c} n! \frac{\prod_{j=1}^{q}(b_j)_n}{\prod_{i=1}^{p}(a_i)_n} = \Gamma(E+1) \frac{\prod_{j=1}^{q}(b_j)_E}{\prod_{i=1}^{p}(a_i)_E} \equiv \rho_{p,q}(E) \quad (3.2)$$

where the Pochhammer symbols are $(x)_E = \dfrac{\Gamma(x+E)}{\Gamma(x)}$.

$$\lim_{d \to c} {}_pF_q\left(\{a\}_1^p; \{b\}_1^q; |z|^2\right) = \lim_{d \to c} \sum_{n=0}^{\infty} \frac{\prod_{i=1}^{p}(a_i)_n}{\prod_{j=1}^{q}(b_j)_n} \frac{(|z|^2)^n}{n!} =$$

$$= \int_0^{\infty} dE\, \frac{\prod_{i=1}^{p}(a_i)_n}{\prod_{j=1}^{q}(b_j)_n} \frac{(|z|^2)^E}{\Gamma(E+1)} = \int_0^{\infty} dE\, \frac{(|z|^2)^E}{\rho_{p,q}(E)} \equiv {}_p\mathcal{F}_q\left(\{a\}_1^p; \{b\}_1^q; |z|^2\right) = \nu_{p,q}(|z|^2) \quad (3.3)$$

It is observed that a function similar to the generalized hypergeometric function is obtained, but defined not by a sum, but by an integral. We will call this new function, *integral generalized hypergeometric function* (int-GHF), and noted by ${}_p\mathcal{F}_q\left(\{a\}_1^p; \{b\}_1^q; |z|^2\right)$. This is just \nu-function. Let's note that the last integral has the same structure as the function \nu $\nu(|z|^2)$, but much more general, if we take into account Eq. (2.12). That is why we will call it the *generalized \nu function* (G\nu) $\nu_{p,q}(|z|^2)$, it being equal even to *integral generalized hypergeometric function* (int-GHF), ${}_p\mathcal{F}_q\left(\{a\}_1^p; \{b\}_1^q; |z|^2\right)$:

$$v_{p,q}(|z|^2) \equiv \int_0^\infty dE \frac{(|z|^2)^E}{\rho_{p,q}(E)} = \int_0^\infty dE \frac{\prod_{i=1}^p (a_i)_E}{\prod_{j=1}^q (b_j)_E} \frac{(|z|^2)^E}{\Gamma(E+1)} \equiv {}_p\mathcal{F}_q(\{a\}_1^p;\{b\}_1^q;|z|^2) \tag{3.4}$$

In the special case, for $p = q = 0$, we have

$$v_{0,0}(|z|^2) \equiv v(|z|^2) = \int_0^\infty dE \frac{(|z|^2)^E}{\rho_{0,0}(E)} = \int_0^\infty dE \frac{(|z|^2)^E}{\Gamma(E+1)} \equiv {}_0\mathcal{F}_0(\ ;\ ;|z|^2) = exp(|z|^2) \tag{3.5}$$

In this context, we can say that the function $v(|z|^2)$ is an *integral exponential function* $exp(|z|^2)$.

*Note:* Do not confuse the *integral exponential function* $exp(|z|^2)$ with the *exponential integral function* $\text{Ei}(x)$, for real $x$, which is a special function on the complex plane, defined as

$$\text{Ei}(x) = -\int_{-x}^\infty dt \frac{e^{-t}}{t} = \int_{-\infty}^x dt \frac{e^t}{t} \tag{3.6}$$

Consequently, the Gd-cL of GH-CSs is

$$\lim_{d \to c}|z> \equiv |z>= \frac{1}{\sqrt{v_{p,q}(|z|^2)}} \int_0^\infty dE \frac{z^E}{\sqrt{\rho_{p,q}(E)}} |E> \tag{3.7}$$

Since it is about the same complex variable $z$, we kept the same notation, as before, for GH-CSs for continuous spectrum $|z>$.

The overlap of two GH-CSs for continuous spectrum is, consequently

$$<z|z'> = \frac{v_{p,q}(z^*z')}{\sqrt{v_{p,q}(|z|^2)}\sqrt{v_{p,q}(|z'|^2)}} \tag{3.8}$$

Using the action of the generalized creation and annihilation operators, in conjunction with the DOOT rules, the projector onto state $|z>$ is

$$|z><z| = \frac{1}{v_{p,q}(|z|^2)} \int_0^\infty dE \frac{z^E}{\sqrt{\rho_{p,q}(E)}} |E> \int_0^\infty dE' \frac{(z^*)^{E'}}{\sqrt{\rho_{p,q}(E')}} <E'| \tag{3.9}$$

Following the same calculation procedure as before, it will be obtained that the measure of integration in the continuous space will be

$$d\mu^c(z) = \frac{d\varphi}{2\pi} d(|z|^2) \frac{\prod_{i=1}^p \Gamma(a_i)}{\prod_{j=1}^q \Gamma(b_j)} v_{p,q}(|z|^2) G_{p,q+1}^{q+1,0}\left(|z|^2 \left| \begin{array}{c} /;\ \{a_i - 1\}_1^p \\ 0, \{b_j - 1\}_1^q;\ / \end{array}\right.\right) \tag{3.10}$$

where we took into account that the angular integral has the value

$$\int_0^{2\pi} \frac{d\varphi}{2\pi} z^E (z^*)^{E'} = (|z|^2)^E \delta(E - E') \tag{3.11}$$

and we also used the fundamental integral, Eq. (2.12), i.e. the general relation for the classical integral to one Meijer's G-function [6] and also the completeness relation for the $|E>$ vectors:

$$\int_0^\infty dE \, | E ><E | = 1 \quad (3.12)$$

In this context it is confirmed that the c-GH-CSs for continuous spectrum can be expanded according to the $|E>$ vectors, in the form

$$|z> = \frac{1}{\sqrt{\nu_{p,q}(|z|^2)}} \int_0^\infty dE \, \frac{z^E}{\sqrt{\rho_{p,q}(E)}} |E> \quad (3.13)$$

Therefore, this integration measure ensures the validity of the closure (completeness) relationship of the unit operator:

$$\int d\mu^c(z) |z><z| = I \quad (3.14)$$

It is observed that the difference between the mathematical expressions for $\nu(|z|^2)$, Eq. (1.4) compared to the one for $\nu_{p,q}(|z|^2)$, Eq. (3.13), consists in the fact that, in the denominator, instead of $\Gamma(E+1)$, it appears $\rho_{p,q}(E)$, and as we previously pointed out, $\nu(|z|^2) \equiv \nu_{0,0}(|z|^2)$.

After applying the discrete-continuous generalized limit $d \to c$, the actions of the creation and annihilation operators on the vacuum state lead to the relations:

$$|E> = \frac{1}{\sqrt{\rho_{p,q}(E)}} (\mathcal{A}_+)^E |0>, \quad <E| = \frac{1}{\sqrt{\rho_{p,q}(E)}} <0| (\mathcal{A}_-)^E \quad (3.15)$$

Generally, the pair of Hermitian operators $\mathcal{A}_-$ and $\mathcal{A}_+$ acts on the vectors $|E>$ in a way that results from the application of the *discrete – continuous limit* $d \to c$:

$$\mathcal{A}_- |E> = \sqrt{\rho_{p,q}(E)} |E-1>, \quad \mathcal{A}_+ |E> = \sqrt{\rho_{p,q}(E+1)} |E+1>,$$
$$\mathcal{A}_+ \mathcal{A}_- |E> = \rho_{p,q}(E) |E>, \quad \#(\mathcal{A}_+\mathcal{A}_-)^E \# |E> = [\rho_{p,q}(E)]^E |E> \quad (3.16)$$

or, equivalently

$$\mathcal{A}_- = \int_0^\infty dE \sqrt{\rho_{p,q}(E)} |E-1><E|, \quad \mathcal{A}_+ = \int_0^\infty dE \sqrt{\rho_{p,q}(E+1)} |E+1><E|,$$
$$\mathcal{A}_+\mathcal{A}_- = \int_0^\infty dE \, \rho_{p,q}(E) |E><E|, \quad \#(\mathcal{A}_+\mathcal{A}_-)^E \# = \int_0^\infty dE [\rho_{p,q}(E)]^E |E><E| \quad (3.17)$$

Similarly, using the DOOT rules, we will obtain the expression for *generalized projector of the vacuum state of continuous spectrum*:

$$\int_0^\infty dE \, |E><E| = |0><0| \int_0^\infty dE \, \frac{\#(\mathcal{A}_+\mathcal{A}_-)^E \#}{\rho_{p,q}(E)} = |0><0| \, \#\nu_{p,q}(\mathcal{A}_+\mathcal{A}_-)\# = 1 \quad (3.18)$$

$$|0><0| = \frac{1}{\#\nu_{p,q}(\mathcal{A}_+\mathcal{A}_-)\#} \quad (3.19)$$

The projector on the state $|z>$ is obtained analogously as for the usual case, using the DOOT rules:

$$|z><z| = \frac{1}{\nu_{p,q}(|z|^2)} \# \int_0^\infty dE \frac{(z\mathcal{A}_+)^E}{\rho_{p,q}(E)} |0><0| \int_0^\infty dE' \frac{(z^*\mathcal{A}_-)^{E'}}{\rho_{p,q}(E')} \# =$$
$$= \frac{1}{\nu_{p,q}(|z|^2)} \# \frac{\nu_{p,q}(z\mathcal{A}_+)\nu_{p,q}(z^*\mathcal{A}_-)}{\nu_{p,q}(\mathcal{A}_+\mathcal{A}_-)} \#$$
(3.20)

where, according to the DOOT rules, the vacuum projector $|0><0|$ can be removed from the sign $\#\ldots\#$.

The completeness relation of the c-GH-CSs lead to

$$\int d\mu^c(z)|z><z| = \frac{\prod_{j=1}^q \Gamma(b_j)}{\prod_{i=1}^p \Gamma(a_i)} \# \frac{1}{\nu_{p,q}(\mathcal{A}_+\mathcal{A}_-)} \# \times$$
$$\times \int_0^\infty d(|z|^2) G_{p,q+1}^{q+1,0}\left(|z|^2 \left| \begin{array}{c} /\,; \quad \{a_i-1\}_1^p \\ 0, \{b_j-1\}_1^q\,; \quad / \end{array}\right.\right) \int_0^{2\pi} \frac{d\varphi}{2\pi} \# \nu_{p,q}(z\mathcal{A}_+)\nu_{p,q}(z^*\mathcal{A}_-) \# = 1$$
(3.21)

because the angular integral is

$$\int_0^{2\pi} \frac{d\varphi}{2\pi} \# \nu_{p,q}(z\mathcal{A}_+)\nu_{p,q}(z^*\mathcal{A}_-) \# = \int_0^\infty dE \frac{\#(\mathcal{A}_+\mathcal{A}_-)^E \#}{[\rho_{p,q}(E)]^2} (|z|^2)^E$$
(3.22)

In addition, from the completeness relation, the following integral in complex space results, which refers to the \nu-function with operatoorial argument:

$$\int \frac{d^2z}{\pi} G_{p,q+1}^{q+1,0}\left(|z|^2 \left| \begin{array}{c} /\,; \quad \{a_i-1\}_1^p \\ 0, \{b_j-1\}_1^q\,; \quad / \end{array}\right.\right) \# \nu_{p,q}(z\mathcal{A}_+)\nu_{p,q}(z^*\mathcal{A}_-) \# =$$
$$= \frac{\prod_{j=1}^q \Gamma(b_j)}{\prod_{i=1}^p \Gamma(a_i)} \# \nu_{p,q}(\mathcal{A}_+\mathcal{A}_-) \#$$
(3.23)

On the other hand, the probability density of the transition from state $|E>$ to state $|z>$ has the expression

$$P_{E;p,q}(|z|^2) \equiv |<z|E>|^2 = \frac{1}{\nu_{p,q}(|z|^2)} \frac{(|z|^2)^E}{\rho_{p,q}(E)}$$
(3.24)

whose *generalized inverse limit* $c \to d$ also leads us to the generalized Poisson distribution, which. If we apply the *inverse* limit, i.e. the *continuous – discrete limit* $c \to d$, we obtain just the usually Poisson probability density function for the discrete case.

## 4. Other interesting properties of the generalized \nu function

To verify the expressions obtained regarding the continuous spectrum, we will apply the reciprocal limit, i.e. the *continuous − discrete limit* $c \to d$ ,. As a result, we will have to obtain the preferred counterpart expressions for the discrete (discontinuous) spectrum. For example:

$$\lim_{c \to d} \nu_{p,q}(|z|^2) \equiv \lim_{c \to d} \int_0^\infty dE \frac{(|z|^2)^E}{\rho_{p,q}(E)} = \sum_{n=0}^\infty \frac{\prod_{i=1}^p (a_i)_n}{\prod_{j=1}^q (b_j)_n} \frac{(|z|^2)^n}{n!} = {}_pF_q(\{a\}_1^p; \{b\}_1^q; |z|^2) \quad (4.1)$$

Particularly, for the case $p = 0$ and $q = 0$, we have ${}_0F_0(;;|z|^2) = e^{|z|^2}$ and we obtain

$$\lim_{c \to d} \nu_{0,0}(|z|^2) \equiv \lim_{c \to d} \nu(|z|^2) = \lim_{c \to d} \int_0^\infty dE \frac{(|z|^2)^E}{\Gamma(E+1)} = \sum_{n=0}^\infty \frac{(|z|^2)^n}{n!} = {}_0F_0(;;|z|^2) = e^{|z|^2} \quad (4.2)$$

We can also define a \nu operator (that is, a \nu function that has an operator as argument):

$$\nu_{p,q}(\mathcal{A}_-) \equiv \int_0^\infty dE \frac{(\mathcal{A}_-)^E}{\rho_{p,q}(E)} \quad , \quad \nu_{p,q}(\mathcal{A}_+) \equiv \int_0^\infty dE \frac{(\mathcal{A}_+)^E}{\rho_{p,q}(E)} \quad (4.3)$$

Its action on c-GH-CSs is easily obtained if we take into account the definition of coherent states in the Barut-Girardello manner [9]:

$$\mathcal{A}_-|z\rangle = z|z\rangle \quad , \quad \langle z|\mathcal{A}_+ = z^*\langle z| \quad , \quad \mathcal{A}_+\mathcal{A}_-|z\rangle = |z|^2|z\rangle \quad (4.4)$$

and we obtain

$$\nu_{p,q}(\mathcal{A}_-)|z\rangle = \int_0^\infty dE \frac{1}{\rho_{p,q}(E)}(\mathcal{A}_-)^E|z\rangle = \int_0^\infty dE \frac{z^E}{\rho_{p,q}(E)}|z\rangle = \nu_{p,q}(z)|z\rangle \quad (4.5)$$

$$\langle z|\nu_{p,q}(\mathcal{A}_+) = \nu_{p,q}(z^*)\langle z| \quad (4.6)$$

Therefore, the function $\nu_{p,q}(z)$ in the variable $z$ is the eigenvalue of the operational function $\nu_{p,q}(\mathcal{A}_-)$ in the coherent state $|z\rangle$.

The mean (expected) value in the coherent state $|z\rangle$ of the ordered product in the sense of DOOT is, then

$$\langle z|\#\nu_{p,q}(\mathcal{A}_+)\nu_{p,q}(\mathcal{A}_-)\#|z\rangle = \nu_{p,q}(z^*)\nu_{p,q}(z) \quad (4.7)$$

It is easy to verify that the action of the \nu-operator (that is, of a \nu-function that has an operator as argument) on c-GH-CSs is

$$\nu_{p,q}(z'^*\mathcal{A}_-)|z'\rangle = \nu_{p,q}(|z'|^2)|z'\rangle \quad , \quad \langle z|\nu_{p,q}(z\mathcal{A}_+) = \nu_{p,q}(|z|^2)\langle z| \quad (4.8)$$

The mean (expected) value in the coherent state $|z\rangle$ is

$$\langle z|\#\nu_{p,q}(z\mathcal{A}_+)\nu_{p,q}(z^*\mathcal{A}_-)\#|z'\rangle = \sqrt{\nu_{p,q}(|z|^2)}\sqrt{\nu_{p,q}(|z'|^2)}\nu_{p,q}(z^*z') \quad (4.9)$$

Let's calculate some (continue!) matrix elements in the c-GH-CSs representation of the function $\nu_{p,q}(...)$ which has different operator functions as its right argument.

$$\langle z|\nu_{p,q}(z'^*\mathcal{A}_-)|z'\rangle = \nu_{p,q}(|z'|^2)\langle z|z'\rangle \quad ,$$
$$(4.10)$$

$$<z|v_{p,q}(z\mathcal{A}_+)|z'> = v_{p,q}(|z|^2)<z|z'>$$

$$<z|v_{p,q}(e^{-z'^*\mathcal{A}_-})|z'> = v_{p,q}(e^{-|z'|^2})<z|z'>$$

$$<z|v_{p,q}(e^{z\mathcal{A}_+})|z'> = v_{p,q}(e^{|z|^2})<z|z'>$$

(4.11)

$$<z|\#v_{p,q}(e^{z\mathcal{A}_+})v_{p,q}(e^{-z'^*\mathcal{A}_-})\#|z'> = v_{p,q}(e^{|z|^2})v_{p,q}(e^{-|z'|^2})<z|z'>$$

$$<z|\#v_{p,q}(e^{z\mathcal{A}_+})v_{p,q}(e^{-z^*\mathcal{A}_-})\#|z> = v_{p,q}(e^{|z|^2})v_{p,q}(e^{-|z|^2})$$

(4.12)

Considering that in the GH-CSs approach, combined with the DOOT rules, the creation and annihilation operators $\mathcal{A}_+$ and $\mathcal{A}_-$ commute, i.e. $[\mathcal{A}_+, \mathcal{A}_-] = 0$, applying the Baker–Campbell–Hausdorff formula

$$\exp(\mathcal{X})\exp(\mathcal{Y}) = \exp(\mathcal{Z})$$

$$\mathcal{Z} = \mathcal{X} + \mathcal{Y} + \frac{1}{2}[\mathcal{X},\mathcal{Y}] + \frac{1}{12}[\mathcal{X},[\mathcal{X},\mathcal{Y}]] + \frac{1}{12}[\mathcal{Y},[\mathcal{X},\mathcal{Y}]] + ...$$

(4.13)

let's calculate the diagonal elements of the function $v_{p,q}(...)$ which has as "argument" *displacement operator*:

$$\mathcal{D}(z) \equiv \# e^{z\mathcal{A}_+ - z^*\mathcal{A}_-}\# = \# e^{z\mathcal{A}_+} e^{-z^*\mathcal{A}_-}\#$$

(4.14)

The final result is (see the deduction in the Appendix)

$$<z|v_{p,q}(\mathcal{D}(z))|z> = <z|v_{p,q}(\# e^{z\mathcal{A}_+ - z^*\mathcal{A}_-}\#)|z> = \int_0^\infty dE \frac{1}{\Gamma(E+1)} = v_{p,q}(1)$$

(4.15)

One of the DOOT rules specifies that, inside the sign #...#, the creation / annihilation operators $\mathcal{A}_+$ and $\mathcal{A}_-$ are considered as simple c-numbers, so they can be removed from under the integral sign [15]. Therefore, we can replace these operators by some numbers (scalar quantities): $\mathcal{A}_+ \to x$ and $\mathcal{A}_- \to y$.

Let's deal with some integrals in which the function is involved $v_{p,q}(...)$.

From the closure (completeness) relationship of the unit operator making the substitutions, we arrive at the two following integrals of fundamental importance in the CSs approach:

- Integral in the complex space, from Eq. (3.23):

$$\int \frac{d^2 z}{\pi} G_{p,q+1}^{q+1,0}\left(|z|^2 \left|\begin{array}{c} /\,; \quad \{a_i - 1\}_1^p \\ 0, \{b_j - 1\}_1^q\,; \quad / \end{array}\right.\right) v_{p,q}(xz) v_{p,q}(yz^*) = \frac{\prod_{j=1}^q \Gamma(b_j)}{\prod_{i=1}^p \Gamma(a_i)} v_{p,q}(xy)$$

(4.16)

- Integral in real space, from Eqs. (3.9), (3.10), and (3.14):

$$\int_0^\infty d(|z|^2)(|z|^2)^E G_{p,q+1}^{q+1,0}\left(|z|^2 \left|\begin{array}{c} /\,; \quad \{a_i - 1\}_1^p \\ 0, \{b_j - 1\}_1^q\,; \quad / \end{array}\right.\right) = \frac{\prod_{j=1}^q \Gamma(b_j)}{\prod_{i=1}^p \Gamma(a_i)} \rho_{p,q}(E)$$

(4.17)

By particularizing the indices $p$ and $q$, the sets of numbers $\{a_i\}_1^p$ and $\{b_j\}_1^q$, as well as by conveniently changing the integration variable, with the help of these integrals we can calculate various other integrals by implicating the $v_{p,q}(x|z|^2)$ function, where $x$ is a real or complex number.

As a first example, we will have

$$\int_0^\infty d(|z|^2) v_{p,q}\left(\frac{1}{x}|z|^2\right) G_{p,q+1}^{q+1,0}\left(|z|^2 \Big| \begin{array}{l} /\,; \quad \{a_i-1\}_1^p \\ 0, \{b_j-1\}_1^q\,; \quad / \end{array}\right) = \frac{\prod_{j=1}^q \Gamma(b_j)}{\prod_{i=1}^p \Gamma(a_i)} \frac{1}{\log x} \quad (4.18)$$

Particularly, if $x=s$, $|z|^2 = st$, $p=0$, $q=0$, $v_{0,0}(...) = v(...)$, we obtain (see, Appendix):

$$\int_0^\infty dt\, e^{-st} v(t) = \frac{1}{s \log s} \quad (4.19)$$

which is in fact the Laplace transform of the $v(t)$ function, the formula that appears in Erdelyi's book [1], page 222.

On the other hand, if $p=1$, $a_1 = 1$, $q=1$, $b_1 = b+1$, in this situation we have

$G_{1,2}^{2,0}\left(|z|^2 \Big| \begin{array}{l} /\,; \quad 0 \\ 0,b\,; \quad / \end{array}\right) = G_{0,1}^{1,0}(|z|^2 | b) = e^{-|z|^2}(|z|^2)^b$, $\rho_{1,1}(E) = \Gamma(b+1+E)$, and the integral becomes

$$\int_0^\infty d(|z|^2) v_{1,1}\left(\frac{1}{x}|z|^2\right) e^{-|z|^2} (|z|^2)^b = \int_0^\infty dE \frac{x^{-E}}{\rho_{1,1}(E)} \int_0^\infty d(|z|^2) e^{-|z|^2} (|z|^2)^{b+E} =$$

$$= \int_0^\infty dE \frac{x^{-E}}{\Gamma(b+1+E)} \Gamma(b+1)\Gamma(b+1+E) = \Gamma(b+1) \int_0^\infty dE\, x^{-E} = \Gamma(b+1) \frac{1}{\log x} \quad (4.20)$$

Generally, if we choose $v_{0,0}(e^{-s|z|^2})$, then, after developing in the power series, we will call on the Laplace transform of the Meijer G-function (see, Appendix):

$$\int_0^\infty d(|z|^2) v_{0,0}(e^{-s|z|^2}) G_{p,q+1}^{q+1,0}\left(|z|^2 \Big| \begin{array}{l} /\,; \quad \{a_i-1\}_1^p \\ 0, \{b_j-1\}_1^q\,; \quad / \end{array}\right) =$$

$$= \sum_{l=0}^\infty \rho_{p,q}(l) \frac{(-1)^l}{l!} s^l \left(\frac{\partial}{\partial s}\right)^l v(e^{-s}) \quad (4.21)$$

A similar integral can be deduced for the function $v(z,\alpha)$ (see, Appendix):

$$\int_0^\infty d(|z|^2) v\left(\frac{1}{C}|z|^2, \alpha\right) G_{p,q+1}^{q+1,0}\left(|z|^2 \Big| \begin{array}{l} /\,; \quad \{a_i-1\}_1^p \\ 0, \{b_j-1\}_1^q\,; \quad / \end{array}\right) =$$

$$= C^{-\alpha} \frac{\prod_{j=1}^q \Gamma(b_j+\alpha)}{\prod_{i=1}^p \Gamma(a_i+\alpha)} v_{p,q+1}(C^{-1}) \quad (4.22)$$

Particularly, in Eq. (4.16), for the case $p = 0$ and $q = 0$, we have $G_{0,1}^{1,0}(|z|^2|0) = e^{-|z|^2}$ and we obtain

$$\int \frac{d^2 z}{\pi} e^{-|z|^2} \nu(x z) \nu(y z^*) = \nu(x y) \tag{4.23}$$

At the end of this section, it is necessary to make the following *observation*: When the generalized function \nu(x) appears under the integral sign, it can also be defined by other structure functions, which have other indices and other sets of numbers, for example:

$$\nu_{r,s}(X) \equiv \int_0^\infty dE \, \frac{X^E}{\rho_{r,s}(E)} \equiv {}_r\mathcal{F}_s(\{c\}_1^r; \{d\}_1^s; X) \quad , \quad \rho_{r,s}(E) = \Gamma(E+1) \frac{\prod_{j=1}^s (d_j)_E}{\prod_{i=1}^r (c_i)_E} \tag{4.24}$$

The only reason why, throughout the paper, we used $\rho_{r,s}(E)$ was to not complicate the formulas too much.

## 5. Concluding remarks

We sought to expand the properties and applications of the \nu-function $\nu(...)$, matters that have been very little (or almost not at all !) treated in the specialized literature, and started from a first application, examined by us in a previous paper [8]. We examine the connection between the formalism of coherent states for continuous spectra and the function $\nu(...)$. There we showed that the function $\nu(...)$ plays the role of normalization function of the coherent states $|z\rangle$ associated with the continuous spectrum of a quantum system. This role could be highlighted through the formulation and use of the *discrete – continuous limit* $d \to c$, whereby any quantity or observable characteristic of the discrete spectrum $X_d(n, n_{\max})$ has a counterpart $X_c(E)$ from the continuous spectrum.

In this paper we generalized the definition for the function $\nu(z)$, in the sense that in the denominator, instead of the gamma function $\Gamma(E+1)$, where $E$ are the eigenvalues of the continuous energy spectrum of the quantum system, a more complicated function appears, containing products and ratios of gamma functions, $\rho_{p,q}(E)$, called the structure function. This structure function, in its *discrete (discontinuous) form*, is directly related to the most general coherent states, i.e. generalized hypergeometric coherent states (GH-CSs). Thus, we introduced a new function - *generalized \nu-function*, $\nu_{p,q}(z)$, and the usual \nu-function is a particular case of it: $\nu(z) = \nu_{0,0}(z)$. We have examined the properties of this new function.

In addition, we also dealt with the generalized functions $\nu_{p,q}(...)$ whose argument depends on the creation or annihilation operators that define the generalized hypergeometric coherent states. The results, involving the $\nu_{p,q}(...)$ functions, are consistent with the well-known relations from the theory of coherent states: $f(\mathcal{A}_-)|z\rangle = f(z)|z\rangle$ and $\langle z| f(\mathcal{A}_+) = \langle z| f(z^*)$. These relations involving operators are important because, due to the rules of the new *diagonal operational ordering technique* (DOOT), the operators are considered to be simple c-numbers, which can be removed from under the DOOT sign # ... # [15]. Consequently, the operators can be replaced by some scalar quantities, thus obtaining new classical mathematical relationships. As applications, we deduced some integrals involving the generalized functions $v_{p,q}(...)$ and we examined some examples. In this manner we opened the way for obtaining several types of integrals and relations involving the functions $v_{p,q}(...)$, by particularizing the indices $p$ and $q$, and the sets of numbers $\{a_i\}_1^p$ and $\{b_j\}_1^q$, and by conveniently changing the integration variables.

In conclusion, let us point out that all the calculations in this paper were made based on the coherent states defined in the Barut-Girardello sense [9]. In principle, the previously presented calculations remain valid also for dual coherent states, defined in the sense of Klauder-Perelomov $|z>_{KP}$ [10]:

$$|z>_{KP} = \frac{1}{\sqrt{{}_qF_p\left(\{b_j\}_1^q; \{a_i\}_1^p; |z|^2\right)}} \exp(z\mathcal{A}_+ - z^*\mathcal{A}_-)|0> =$$

$$= \frac{1}{\sqrt{{}_qF_p\left(\{b_j\}_1^q; \{a_i\}_1^p; |z|^2\right)}} \sum_{n=0}^{\infty} \frac{z^n}{\sqrt{\rho_{q,p}(n)}} |n>,$$

(5.1)

$$\rho_{q,p}(n) = \Gamma(n+1) \frac{\prod_{i=1}^{p}(a_i)_n}{\prod_{j=1}^{q}(b_j)_n}$$

By applying the *discrete – continuous limit* $d \to c$ the same results are obtained, taking into account the duality between the two types of coherent states, BG-CSs versus KP-CSs [11].

**Appendix**

**1. Deriving the formula (4.21):**

$$\int_0^\infty d(|z|^2) v_{0,0}\left(e^{-s|z|^2}\right) G_{p,q+1}^{q+1,0}\left(|z|^2 \left| \begin{array}{c} /; \quad \{a_i-1\}_1^p \\ 0, \{b_j-1\}_1^q; \quad / \end{array}\right.\right) =$$

$$= \int_0^\infty dE \frac{(e^{-s})^E}{\Gamma(E+1)} \sum_{l=0}^{\infty} \frac{(-sE)^l}{l!} \int_0^\infty d(|z|^2)(|z|^2)^l G_{p,q+1}^{q+1,0}\left(|z|^2 \left| \begin{array}{c} /; \quad \{a_i-1\}_1^p \\ 0, \{b_j-1\}_1^q; \quad / \end{array}\right.\right) =$$

$$= \frac{\prod_{j=1}^{q}\Gamma(b_j)}{\prod_{i=1}^{p}\Gamma(a_i)} \sum_{l=0}^{\infty} \frac{s^l}{l!} (1)_l \frac{\prod_{j=1}^{q}(b_j)_l}{\prod_{i=1}^{p}(a_i)_l} \int_0^\infty dE \frac{(e^{-s})^E}{\Gamma(E+1)} (-E)^l =$$

(A 4.21)

$$= \sum_{l=0}^{\infty} \frac{s^l}{l!} (1)_l \frac{\prod_{j=1}^{q}\Gamma(b_j+l)}{\prod_{i=1}^{p}\Gamma(a_i+l)} (-1)^l \left(\frac{\partial}{\partial s}\right)^l \int_0^\infty dE \frac{(e^{-s})^E}{\Gamma(E+1)} = \sum_{l=0}^{\infty} \rho_{p,q}(l) \frac{(-1)^l}{l!} s^l \left(\frac{\partial}{\partial s}\right)^l v(e^{-s})$$

**2. Deriving the formula (4.15):**

$$<z|v_{p,q}(\mathcal{D}(z))|z> = <z|v_{p,q}\left(\#e^{z\mathcal{A}_+-z^*\mathcal{A}_-}\#\right)|z> = <z|\int_0^\infty dE \frac{\#\left(e^{z\mathcal{A}_+-z^*\mathcal{A}_-}\right)^E\#}{\rho_{p,q}(E)}|z> =$$

$$= \int_0^\infty dE \frac{1}{\rho_{p,q}(E)} <z|\#\left(e^{z\mathcal{A}_+-z^*\mathcal{A}_-}\right)^E\#|z> = \int_0^\infty dE \frac{1}{\rho_{p,q}(E)} <z|\#\left(e^{z\mathcal{A}_+}e^{-z^*\mathcal{A}_-}\right)^E\#|z> =$$

$$= \int_0^\infty dE \frac{1}{\rho_{p,q}(E)} \#<z|\#\left(e^{z\mathcal{A}_+E}\right)\left(e^{-z^*\mathcal{A}_-E}\right)\#|z> =$$

$$= \int_0^\infty dE \frac{1}{\rho_{p,q}(E)} \#\left[\sum_{j=0}^\infty \frac{(zE)^j}{j!}<z|(\mathcal{A}_+)^j\right]\left[\sum_{l=0}^\infty \frac{(z^*E)^l}{l!}(\mathcal{A}_-)^l|z>\right]\# = \quad (A\ 4.15)$$

$$= \int_0^\infty dE \frac{1}{\rho_{p,q}(E)} \left[\sum_{j=0}^\infty \frac{(zE)^j}{j!}<z|(z^*)^j\right]\left[\sum_{l=0}^\infty \frac{(z^*E)^l}{l!}z^l|z>\right] =$$

$$= \int_0^\infty dE \frac{1}{\rho_{p,q}(E)} \left[\sum_{j=0}^\infty \frac{(|z|^2E)^j}{j!}\right]\left[\sum_{l=0}^\infty \frac{(|z|^2E)^l}{l!}\right] = \int_0^\infty dE \frac{1}{\rho_{p,q}(E)}\left(e^{|z|^2}\right)\left(e^{-|z|^2}\right) =$$

$$= \int_0^\infty dE \frac{1}{\rho_{p,q}(E)} = v_{p,q}(1)$$

**3. Deriving the formula (4.18):**

$$\int_0^\infty d(|z|^2) v_{p,q}\left(\frac{1}{x}|z|^2\right) G_{p,q+1}^{q+1,0}\left(|z|^2 \left| \begin{array}{c} /; \quad \{a_i-1\}_1^p \\ 0,\{b_j-1\}_1^q;\quad / \end{array}\right.\right) =$$

$$= \int_0^\infty dE \frac{x^{-E}}{\rho_{p,q}(E)} \int_0^\infty d(|z|^2)(|z|^2)^E G_{p,q+1}^{q+1,0}\left(|z|^2 \left| \begin{array}{c} /; \quad \{a_i-1\}_1^p \\ 0,\{b_j-1\}_1^q;\quad / \end{array}\right.\right) =$$

$$= \frac{\prod_{j=1}^q \Gamma(b_j)}{\prod_{i=1}^p \Gamma(a_i)} \int_0^\infty dE \frac{x^{-E}}{\rho_{p,q}(E)} \rho_{p,q}(E) = \frac{\prod_{j=1}^q \Gamma(b_j)}{\prod_{i=1}^p \Gamma(a_i)} \int_0^\infty dE\, x^{-E} = \frac{\prod_{j=1}^q \Gamma(b_j)}{\prod_{i=1}^p \Gamma(a_i)} \int_0^\infty dE\, e^{-E\log x} = \quad (A\ 4.18)$$

$$= \frac{\prod_{j=1}^q \Gamma(b_j)}{\prod_{i=1}^p \Gamma(a_i)} (-1)\frac{1}{\log x} x^{-E}\bigg|_0^\infty = \frac{\prod_{j=1}^q \Gamma(b_j)}{\prod_{i=1}^p \Gamma(a_i)} \frac{1}{\log x}$$

**4. Deriving the formula (4.22):**

$$\text{Int} \equiv \int_0^\infty d(|z|^2) \nu\left(\frac{1}{C}|z|^2, \alpha\right) G_{p,q+1}^{q+1,0}\left(|z|^2 \left| \begin{array}{c} /\,; \quad \{a_i-1\}_1^p \\ 0, \{b_j-1\}_1^q\,; \quad / \end{array}\right.\right) =$$

$$= \int_0^\infty dE \frac{C^{-(\alpha+E)}}{\Gamma(\alpha+E+1)} \int_0^\infty d(|z|^2) (|z|^2)^{\alpha+E} G_{p,q+1}^{q+1,0}\left(|z|^2 \left| \begin{array}{c} /\,; \quad \{a_i-1\}_1^p \\ 0, \{b_j-1\}_1^q\,; \quad / \end{array}\right.\right) = \quad \text{(A 4.22/1)}$$

$$= \int_0^\infty dE \frac{C^{-(\alpha+E)}}{\Gamma(\alpha+E+1)} \Gamma(\alpha+E+1) \frac{\prod_{j=1}^q \Gamma(b_j+\alpha+E)}{\prod_{i=1}^p \Gamma(a_i+\alpha+E)} = \int_0^\infty dE \frac{C^{-(\alpha+E)}}{\Gamma(\alpha+E+1)} \rho_{p,q}(\alpha+E)$$

where

$$\rho_{p,q}(\alpha+E) = \Gamma(\alpha+E+1) \frac{\prod_{j=1}^q \Gamma(b_j+\alpha) \prod_{j=1}^q (b_j+\alpha)_E}{\prod_{i=1}^p \Gamma(a_i+\alpha) \prod_{i=1}^p (a_i+\alpha)_E} \quad \text{(A 4.22/2)}$$

$$\text{Int} \equiv C^{-\alpha} \frac{\prod_{j=1}^q \Gamma(b_j+\alpha)}{\prod_{i=1}^p \Gamma(a_i+\alpha)} \int_0^\infty dE\, (1)_E \frac{\prod_{j=1}^q (b_j+\alpha)_E}{\prod_{i=1}^p (a_i+\alpha)_E} \frac{C^{-E}}{\Gamma(E+1)} =$$

$$= C^{-\alpha} \frac{\prod_{j=1}^q \Gamma(b_j+\alpha)}{\prod_{i=1}^p \Gamma(a_i+\alpha)} \nu_{p,q+1}(C^{-1}) \quad \text{(A 4.22/3)}$$